\documentstyle[12pt]{article}
\newcommand{\bea}{\begin{eqnarray}}
\newcommand{\eea}{\end{eqnarray}}

\newcommand{\bra}{(0|}
\newcommand{\ket}{|0)}
\newcommand{\tr}{\mathop {\rm tr}}
\newcommand{\sign}{\mathop {\rm sign}}
\newcommand{\eps}{\varepsilon}

\newcommand{\be}[1]{\begin{equation}\label{#1}}
\newcommand{\ba}[1]{\begin{eqnarray}\label{#1}}
\newcommand{\ee}{\end{equation}}
\newcommand{\ea}{\end{eqnarray}}
\newcommand{\non}{\nonumber\\\rule{0pt}{30pt}}
\newcommand{\nona}[1]{\nonumber\\\rule{0pt}{#1pt}}
\newcommand{\num}{\\\rule{0pt}{20pt}}

\newcommand{\dis}{\displaystyle}
\newcommand{\Eq}[1]{(\ref{#1})}
\makeatletter
\@addtoreset{equation}{section}
\makeatother

\newtheorem{thm}{Theorem}[section]
\begin{document}
%
%

\begin{flushright}
September 1997 \\
\end{flushright}

\begin{center}
{\large\bf Normal Ordering in the Theory of Correlation Functions
of Exactly Solvable Models}

\vspace{2mm}
{  Vladimir Korepin\footnote{E-mail:korepin@insti.physics.sunysb.edu}}
and
{\normalsize Nikita Slavnov\footnote{E-mail:nslavnov@mi.ras.ru}}\\
\vskip .5em
\vspace{1cm}
\raisebox{2mm}{\footnotesize 1} 
{\it Institute for Theoretical Physics,\\ State
University of New York at Stony Brook, NY 11794-3840, 
USA\\} 
\vskip .4cm 
\raisebox{2mm}{\footnotesize  2} {\it Steklov 
Mathematical Institute,\\
Gubkina 8, 117966,  Moscow, Russia\\}

\end{center}

\vskip4pt

\begin{abstract}
\noindent
We study models of quantum statistical mechanics which can be solved
by the algebraic Bethe ansatz.  The general method of calculation of
correlation functions is based on the method of determinant
representations.  The auxiliary Fock space and auxiliary Bose fields are
introduced in order to remove the two body scattering and represent
correlation functions as a mean value of a determinant of a Fredholm
integral operator; the representation has a simple form for large
space and time separations.  In this paper we explain how to calculate
the mean value in the auxiliary Fock space of asymptotic expression of
the Fredholm determinant.  It is necessary for the evaluation of the
asymptotic form of the physical correlation functions.
\end{abstract}
\vfill
\newpage
\section{Introduction}\label{intro}

In this paper we use an example of the quantum nonlinear
Schr\"odinger equation, i.e. the one-dimensional  Bose gas with 
delta-function interactions, in order to illustrate the development in the 
theory of quantum correlation functions.  The correlation function 
of local fields in this model was studied in previous papers 
\cite{ref2}--\cite{ref5}, and its determinant representation was 
obtained in the paper \cite{ref2}. The representation of the correlation
function in terms of the Fredholm determinant of a linear integral operator
is the basis of our approach. The differential equations for the
Fredholm determinant had been obtained in \cite{ref3}.  They are directly 
related to the classical nonlinear Schr\"odinger equation.
These differential equation were solved in the asymptotic 
regime of large space and time separations;  the 
simplified asymptotic form of Fredholm determinant was 
obtained in \cite{ref4}--\cite{ref5}.  This expression is an operator 
in an auxiliary Fock space. In order to find the asymptotics of the
correlation function, one should calculate  the vacuum mean value of this 
expression.  It is a necessary step in the calculation of asymptotics
of physical correlation functions.  The problem of evaluation of the
vacuum expectation (mean value) is a combinatorial problem, closely
related to the procedure of the normal ordering in the quantum field
theory. In the present paper we are studying just this problem.

We remind in brief the basic definitions of the model under 
consideration for the reader's convenience. The quantum nonlinear 
 Schr\"odinger equation can be described in terms of the canonical 
 Bose fields $\psi(x,t)$ and $\psi^\dagger(x,t)$ ($x\in R$) obeying 
 the equal time commutation relations 
\be{cancom}
\left[ 
\psi(x,t),\psi^\dagger(y,t)\right] = \delta(x-y).  
\ee 
The Hamiltonian and momentum of the model are 
\be{Ham}
 H=\int dx \Bigl( 
\partial_x \psi^\dagger(x) \partial_x\psi(x) + c \psi^\dagger(x) 
 \psi^\dagger(x) \psi(x) \psi(x) - h \psi^\dagger(x) \psi(x)  \Bigr), 
\ee
and
\be{Mom}
P = -i \int dx  ~\psi^\dagger(x) \partial_x\psi(x) .
\ee
Here $0<c\leq\infty$ is the coupling constant and $h$ is the
chemical potential.  The spectrum of the model was first described by
E.~H.~Lieb and W.~Liniger \cite{ref6}--\cite{ref7}.  The Lax
representation for the
corresponding classical equation of motion 
\be{Lax}
i{\partial\over\partial t}\psi=\left[\psi,H\right]
= -{\partial^2\over\partial x^2}\psi + 2c\psi^\dagger \psi\psi
  - h\psi  ,
\ee
was found by V.~E.~Zakharov and A.~B.~Shabat \cite{ref8}.
The quantum inverse scattering method for the model was formulated by
L.~D.~Faddeev and E.~K.~Sklyanin \cite{ref9}.  

The quantum nonlinear Schr\"odinger equation is equivalent to the 
Bose gas with delta-function interactions.  In the sector with $N$ 
particles the Hamiltonian of the Bose gas is given by 
\be{Ham1}
{\cal H}_N = -\sum_{j=1}^N\frac{\partial^2}{\partial z_j^2} + 
  2c\sum_{1\leq j<k\leq N} \delta(z_j-z_k) - Nh  .  
\ee 
In this paper we shall consider the thermodynamics of this model.  The 
partition function and the free energy of the model are defined by 
\be{Part}
 Z=\tr e^{-\frac{H}{T}} = e^{-\frac{F}{ T}}  .
\ee
The free energy $F$ has been
explicitly represented in terms of the Yang-Yang equation
\cite{ref10}
\be{YY}
\varepsilon(\lambda) = \lambda^2 - h - \frac{T}{2\pi}
\int^\infty_{-\infty} d\mu ~  \frac{2c}{c^2+(\lambda-\mu)^2}
\ln\Bigl(1+ e^{-\frac{\eps(\mu)}{T}} \Bigr),
\ee
\be{Free}
 F=-\frac{T}{
2\pi} \int^\infty_{-\infty} \ln\Bigl(1+ e^{-\frac{\eps(\mu)}{T}}
  \Bigr)   .
\ee
The correlation function,  studied in this paper is defined by
\be{corr}
\langle\psi(0,0) \psi^\dagger(x,t)\rangle_T =
  \frac{\tr \Bigl( e^{-\frac{H}{T}}
  \psi(0,0)\psi^\dagger(x,t) \Bigr)}{
 \tr \Bigl( e^{-\frac{H}{T}} \Bigr)}  .
\ee

The plan of the paper is the following.
In Section 2 we shall remind the reader of the asymptotic expression
for the Fredholm determinant that represents the correlation function.
  We
shall also describe its dependence on the quantum fields.  Section 3
is devoted to the main results of the paper.  In it we develop a
technique for the evaluation of the mean values in auxiliary Fock
space.  It is related to the problems of the normal ordering in quantum
field theory.  In Section 4 we use this technique in order to 
calculate the mean value of the asymptotic expressions for the 
Fredholm determinant.  
\section{Asymptotics of the 
\hspace{-1pt} 
Fredholm Deter\-mi\-nant}

In order to find the determinant representation of the
correlation functions one has to introduce an auxiliary Fock space
and three Bose fields $\psi(\lambda)$,
$\phi(\lambda)$, and $\Phi(\lambda)$, which are linear
combinations of annihilation and creation operators $p(\lambda)$ and
$q(\lambda)$:
\be{repdualfields}
\begin{array}{l}
{\dis
\psi(\lambda)=q_\psi(\lambda)+p_\psi(\lambda),}\nona{15}
{\dis
\phi(\lambda)=q_\phi(\lambda)+p_\phi(\lambda),}\nona{15}
{\dis
\Phi(\lambda)=q_\Phi(\lambda)+p_\Phi(\lambda).}\nona{15}
\end{array}
\ee
The operators $p(\lambda)$ annihilate the Fock vacuum
\be{p}
p(\lambda)\ket=0,
\ee
and the corresponding creation operators $q(\lambda)$
annihilate the dual vacuum
\be{q}
\bra q(\lambda)=0.
\ee
We shall also use the function
\be{h}
h(\lambda,\mu)=(\lambda-\mu+ic)/ic,
\ee
which enters into the non vanishing commutators
\ba{commutators1}
&&\hspace{-1.5cm}{\dis [p_\psi(\lambda),q_\phi(\mu)]
=-[p_\phi(\lambda),q_\psi(\mu)]
  =\ln\left( {h(\mu,\lambda)\over h(\lambda,\mu)}\right),
}\num
&&\hspace{-1.5cm}{\dis
[p_\psi(\lambda),q_\Phi(\mu)] =
[p_\Phi(\lambda),q_\psi(\mu)] = [p_\psi(\lambda),q_\psi(\mu)] 
 =\ln\biggl(  h(\lambda,\mu) h(\mu,\lambda) \biggr).}
\label{commutators2}
\ea
The relation of these quantum fields to the ones $\phi_{A_2}$,
$\phi_{D_1}$ used in the papers \cite{ref2}-\cite{ref5} are
\be{Arel}
\phi(\lambda)=\phi_{A_2}(\lambda) - \phi_{D_1}(\lambda),
\qquad
\Phi(\lambda)=\phi_{A_2}(\lambda) + \phi_{D_1}(\lambda).
\ee
The vacuum vector is normalized by unity $\bra 0)=1$.

The quantum fields \Eq{repdualfields}  are linear combinations of the
three canonical Bose fields.  The derivative of the field
$\psi(\lambda)$ also will be important:
\be{Aderpsi}
\psi'(\lambda)\equiv
{\partial\over\partial\lambda} \psi(\lambda) = q'_\psi(\lambda) +
 p'_\psi(\lambda).
\ee
Nonzero commutation relations between 
the derivatives of annihilation operators $ p(\lambda)$
and creation operators $q(\lambda)$ are:
\ba{dircom1}
&&\hspace{-2cm}{\dis
[p'_\psi(\lambda),q_\phi(\mu)]
=[p_\phi(\lambda),q'_\psi(\mu)]
 = \frac{2ic}{ c^2+(\lambda-\mu)^2},}\num
&&\hspace{-2cm}{\dis \label{dircom2}
[p'_\psi(\lambda),q'_\psi(\mu)]
= \left(\frac{1}{\lambda-\mu+ic}\right)^2 +
 \left(\frac{1}{\mu-\lambda+ic}\right)^2, }\num
&&\hspace{-2cm}{\dis \label{dircom3}
[p_\Phi(\lambda),q'_\psi(\mu)] = [p_\psi(\lambda),q'_\psi(\mu)] =
-[p'_\psi(\lambda),q_\Phi(\mu)]=
 \frac{2(\mu-\lambda)}{ (\lambda-\mu)^2+c^2}.}
\ea

It is worth mentioning that quantum fields \Eq{repdualfields}
belong to the same Abelian sub-algebra.  They all commute:
\be{abel}
\begin{array}{l}
{\dis
[\psi(\lambda),\psi(\mu)] = [\psi(\lambda),\phi(\mu)]
 = [\psi(\lambda),\Phi(\mu)] = 0,}\non
{\dis
[\phi(\lambda),\Phi(\mu)] = [\phi(\lambda),\phi(\mu)]
 = [\Phi(\lambda),\Phi(\mu)] = 0.}
\end{array}
\ee
This property plays a very important role for the calculation
of vacuum mean values in auxiliary Fock space.

In the paper \cite{ref2} the correlation function of local fields of 
the quantum  nonlinear Schr\"odinger equation was represented as a mean 
value of a determinant of an integral operator, 
depending on the fields \Eq{repdualfields}.  At large space $x$ and 
time $t$ separation the determinant simplifies  to \cite{ref5}:
\be{asy01}
\langle\psi(0,0)\psi^\dagger(x,t)\rangle_T =
\bra Q(x,t)\left[1+o\left(\frac{1}{\sqrt t}\right)\right]\ket,
\ee
where  $Q(x,t)$ is an operator in auxiliary Fock space
\pagebreak[3]
\ba{asy1}
&&{\dis
Q(x,t)=C([\phi(\lambda)],[\Phi(\lambda)])
(2t)^{(\nu-1)^2/2}e^{\psi(\lambda_0)-it\lambda_0^2 -iht}
}\non
&&{\dis \times
\exp\left\{ \frac{1}{2\pi} \int_{-\infty}^\infty d\lambda~  \bigl(
 \vert x-2\lambda t\vert -i~\sign(\lambda-\lambda_0)
\psi'(\lambda) \bigr) \right.}\non
&&{\dis
\left.\vphantom{\int_\infty^\infty}\hspace{1.5cm} \times
\ln\left[1-\vartheta(\lambda) \biggl(1+e^{\phi(\lambda)~\sign
(\lambda-\lambda_0)}\biggr) \right]\right\} .}
\ea
Here $C$ is a smooth bounded functional, which may depend on $x$ and $t$
only through the ratio $x/2t=\lambda_0$, which remains fixed. The 
Fermi weight $\vartheta(\lambda)$ is defined by 
\be{AFermi}
\vartheta(\lambda)=\Bigl(1+e^{\eps(\lambda)/T}\Bigr)^{-1},
\ee
and
\be{nu}
\nu=\frac{i}{ 2\pi} \ln\left\{\Bigl[ 1-\vartheta(\lambda_0)
(1+e^{-\phi(\lambda_0)})\Bigr]
\Bigl[1-\vartheta(\lambda_0)(1+e^{\phi(\lambda_0)})\Bigr]\right\}.
\ee

In this paper our main purpose is to evaluate the mean value of the right
hand side of \Eq{asy1}. Due to the relations \Eq{commutators1} and 
\Eq{commutators2}, the creation and annihilation parts of the fields
$\phi$ and  $\Phi$ commute with each other. Thus the only nonzero
contribution to the vacuum mean value is provided by normal ordering
of expressions containing the field $\psi(\lambda)$. It is easy to
find the contribution of the factor $e^{\psi(\lambda_0)}$ (see
\Eq{asy1}). Indeed let us move this exponent to the left
\be{Aleftmove}
\bra e^{\psi(\lambda_0)} =\bra e^{p_\psi(\lambda_0)}.
\ee
After this one can move
$e^{p_\psi(\lambda_0)}$ to the right using obvious relations:
\be{Alm1}
[p_\psi(\lambda_0),\phi(\lambda_0)]=0,
\ee
\be{lm2}
[p_\psi(\lambda_0),\nu]=0,
\ee
\be{lm3}
e^{p_\psi(\lambda_0)} \phi(\mu) = \left( \phi(\mu) +
  \ln\frac{h(\mu,\lambda_0)}{ h(\lambda_0,\mu)}\right)
e^{p_\psi(\lambda_0)},
\ee
\be{lm4}
e^{p_\psi(\lambda_0)} \Phi(\mu) = \Bigl( \Phi(\mu) +
\ln[h(\lambda_0,\mu) h(\mu,\lambda_0)]\Bigr) e^{p_\psi(\lambda_0)},
\ee
\be{lm5}
e^{p_\psi(\lambda_0)} \psi'(\mu) = \left( \psi'(\mu) +
\frac{2(\mu-\lambda_0)}{(\mu-\lambda_0)^2 + c^2} \right)
e^{p_\psi(\lambda_0)},
\ee
and
\be{lm6}
e^{p_\psi(\lambda_0)} \ket=\ket.
\ee
Thus we arrive at
\ba{asy2}
&&{\dis
\bra Q(x,t)\ket=
\bra
\widetilde C([\phi(\lambda)],[\Phi(\lambda)])
 (2t)^{(\nu-1)^2/2} e^{-it\lambda_0^2 -iht}
}\non
&&{\dis
  \hskip -.4in\times
\exp\left\{ {1\over 2\pi} \int_{-\infty}^\infty d\lambda~  \biggl(
 \vert x-2\lambda t\vert -i~\sign(\lambda-\lambda_0)
\Bigl[\psi'(\lambda)+\frac{2(\lambda-\lambda_0)}
{(\lambda-\lambda_0)^2+c^2}\Bigr] \biggr)
\right.}\non
&&{\dis
\left. \hspace{-5mm}\times
\ln\left[1-\vartheta(\lambda) \biggl(1+
\exp\Bigl[\sign (\lambda-\lambda_0)
\Bigl(\phi(\lambda)+\ln\frac{h(\lambda,\lambda_0)}
{h(\lambda_0,\lambda)}\Bigr)\Bigr]
\biggr) \right]\right\} \ket.}
\ea
Here the functional $\widetilde C([\phi],[\Phi])$ can be obtained
from the functional $C([\phi],[\Phi])$ by shifting of the arguments
of the last one according to  rules \Eq{lm3} and \Eq{lm4}.

Now the right  hand side of \Eq{asy2} can be written in the following form
\be{form}
\bra Q(x,t)\ket=(0\vert \exp\left\{ \int^\infty_{-\infty} d\lambda
 \psi'(\lambda) f\bigl(\lambda\vert\phi(\lambda)\bigr) \right\}
F([\phi],[\Phi]) \vert 0).
\ee
Here
\ba{Af}
&&{\dis \hspace{-1cm}
f\Bigl(\lambda|\phi(\lambda)\Bigr)=
\frac{\sign(\lambda-\lambda_0)}{2\pi i}}\non
&&{\dis
\times\ln\left[1-\vartheta(\lambda) \biggl(1 +
\exp\Bigl[\sign (\lambda-\lambda_0)
\Bigl(\phi(\lambda)+\ln\frac{h(\lambda,\lambda_0)}
{h(\lambda_0,\lambda)}\Bigr)\Bigr]
\biggr) \right],}
\ea
and
\ba{AF}
&&{\dis\hspace{-2cm}
F([\phi],[\Phi])=\widetilde C([\phi(\lambda)],[\Phi(\lambda)])
 (2t)^{(\nu-1)^2/2} e^{-it\lambda_0^2 -iht}
}\non
&&{\dis
  \hskip -.4in\times
\exp\left\{ {1\over 2\pi} \int_{-\infty}^\infty d\lambda~  \biggl(
 \vert x-2\lambda t\vert -2i~\sign(\lambda-\lambda_0)
\frac{\lambda-\lambda_0}
{(\lambda-\lambda_0)^2+c^2} \biggr)
\right.}\non
&&{\dis
\left. \hspace{-5mm}\times
\ln\left[1-\vartheta(\lambda) \biggl(1+
\exp\Bigl[\sign (\lambda-\lambda_0)
\Bigl(\phi(\lambda)+\ln\frac{h(\lambda,\lambda_0)}
{h(\lambda_0,\lambda)}\Bigr)\Bigr]
\biggr) \right]\right\}. }
\ea
In the next section we shall evaluate the right hand side of \Eq{form}.
\section{Evaluation of the Mean Value}

The main purpose of this section is the evaluation of the mean
value
\be{goal}
(0\vert \exp\left\{ \int^\infty_{-\infty} d\lambda
 \psi'(\lambda) f\bigl(\lambda\vert\phi(\lambda)\bigr) \right\}
F([\phi],[\Phi]) \vert 0).
\ee
Here complex function $f$ becomes an operator,  because it depends
on quantum field $\phi(\lambda)$.
It is worth mentioned that the particular case of \Eq{goal}, when 
$f(\lambda|\phi(\lambda))$ is a linear function of the field 
$\phi(\lambda)$, was first considered in \cite{EFIK}. 

We remind the reader of the definitions
\be{def}
\begin{array}{c}
\psi'(\lambda) = p'_\psi(\lambda) + q'_\psi(\lambda),
\nona{15}
\phi(\lambda) = p_\phi(\lambda) + q_\phi(\lambda),
\nona{15}
\Phi(\lambda) = p_\Phi(\lambda) + q_\Phi(\lambda).
\end{array}
\ee
As usual, the relations $p(\lambda)\vert 0)=0$ and
$(0\vert q(\lambda)=0$ for all $p$ and $q$ are
satisfied as well as the commutation relations
\be{com1}
[p'_\psi(\lambda),q_\phi(\mu)] = \xi(\lambda,\mu) =
  [p_\phi(\lambda),q'_\psi(\mu)],
\ee
\be{com2}
[p'_\psi(\lambda),q_\Phi(\mu)] = {\tilde\xi}(\lambda,\mu)
=[p_\Phi(\lambda),q'_\psi(\mu)], 
\ee
\be{com3}
[p'_\psi(\lambda),q'_\psi(\mu)] = \eta(\lambda,\mu).
\vphantom{\tilde\xi}
\ee
The complex functions $\xi(\lambda,\mu)$ and ${\tilde\xi}(\lambda,\mu)$
are equal to
\be{comfunxi}
\xi(\lambda,\mu)=\xi(\mu,\lambda)={2ic\over c^2+(\lambda-\mu)^2},
\quad\quad
{\tilde\xi}(\lambda,\mu) = {2(\lambda-\mu)\over (\lambda-\mu)^2+c^2},
\ee
and
\be{comfuneta}
\eta(\lambda,\mu)= \left({1\over \lambda-\mu+ic}\right)^2
 + \left({1\over \lambda-\mu-ic}\right)^2   \ .
\ee
However we do not use explicit expressions \Eq{comfunxi} and
\Eq{comfuneta} in this section.

We evaluate the mean value \Eq{goal} in three steps. First, we consider
an auxiliary problem---the scalar case.

\subsection{Scalar Case}

Consider the following mean value
\be{sc1}
(0\vert e^{\left\{ \alpha\psi'(\lambda) 
f\bigl( \phi(\mu)\bigr) \right\} }
 F([\phi],[\Phi]) \vert 0).
\ee
Here $F([\phi],[\Phi])$ is a smooth functional, depending on the fields
$\phi$ and $\Phi$. A complex function $f(z)$ becomes
an operator-valued function while its argument $\phi(\mu)$ is an 
operator.  As usual such an expression should be understood as a 
formal Taylor series; therefore, we assume that $f(z)$ is holomorphic 
within some circle $|z|<\rho_0$.

The parameter $\alpha$ is a complex number such that the following
restriction holds:
\be{restrict}
|\alpha|\cdot\biggl|f\Bigl(z\xi(\lambda,\mu)\Bigr)\biggr|
< |z|,\qquad\mbox{for}\qquad 
|z|=\rho<\frac{\rho_0}{|\xi(\lambda,\mu)|}.  
\ee 
The complex numbers $\lambda$ and $\mu$ are fixed.

Let us decompose the exponent in a Taylor series, 
\be{sc2} 
e^{\left\{ \alpha\psi'(\lambda) f\bigl( \phi(\mu)\bigr) \right\} } =
\sum_{n=0}^\infty \frac{\alpha^n}{n!}
\Bigl( \psi'(\lambda)\Bigr)^n
f^n(\phi(\mu)) .
\ee
We would like to emphasize that here we essentially used the
commutativity of the fields $\psi'$ and $\phi$.

In order to calculate $(0\vert (\psi'(\lambda))^n$ we use the
Cauchy integral representation
\be{intrep}
\bigl(\psi'(\lambda)\bigr)^n =
\left.\frac{d^n}{ dz^n} e^{z\psi'(\lambda)} \right|_{z=0}
 = \frac{n!}{2\pi i} \int\limits_{|z|=\rho} \,dz
\frac{e^{z\psi'(\lambda)} }{z^{n+1}}.
\ee 
The evaluation of $(0\vert e^{z\psi'(\lambda)}$ is a standard 
problem in quantum field theory
\be{sc3} 
(0\vert e^{z\psi'(\lambda)} = 
e^{z^2\eta(\lambda,\lambda)/2} (0\vert 
e^{zp'_\psi(\lambda)}.  
\ee 
After substituting this into (\ref{intrep}), we obtain 
\be{subintrep} 
(0\vert \bigl( \psi'(\lambda)\bigr)^n 
 = \frac{n!}{2\pi i} \int\limits_{|z|=\rho} \,dz
\frac{e^{z^2\eta(\lambda,\lambda)/2} }{z^{n+1}}
 (0 \vert e^{zp'_\psi(\lambda)}. 
\ee
Further substitution into \Eq{sc1} and \Eq{sc2} gives the expression
for the mean value
\ba{interm1}
&&{\dis\hspace{-15mm}
 (0\vert e^{\left\{\alpha \psi'(\lambda)
f\bigl(\phi(\mu)\bigr)  \right\}} F([\phi],[\Phi]) \vert 0)}\non
&&{\dis  =
\sum_{n=0}^\infty
\frac{\alpha^n}{2\pi i} \int\limits_{|z|=\rho} \,dz
\frac{e^{z^2\eta(\lambda,\lambda)/2} }{z^{n+1}}
 (0\vert e^{zp'_\psi(\lambda)}
 f^n(\phi(\mu)) F([\phi],[\Phi]) \vert 0) .}
\ea
Recall that creation and annihilation parts of the fields
$\phi$ and $\Phi$ commute with each other; therefore, we have
\ba{repfiqor}
{\dis (0\vert e^{zp'_\psi(\lambda)} f^n(\phi(\mu)) 
F([\phi],[\Phi]) \vert 0)} &=&{\dis
(0\vert e^{zp'_\psi(\lambda)} f^n(q_\phi(\mu)) 
F([q_\phi],[q_\Phi]) \vert 0) }\non
&&{\dis\hspace{-1.5cm}=
f^n\Bigl(z\xi(\lambda,\mu)\Bigr) (0\vert e^{zp'_\psi(\lambda)}
 F([\phi],[\Phi]) \vert 0).}
\ea 
If we substitute this into (\ref{interm1}) then we may calculate the
sum with respect to $n$. Due to the restriction \Eq{restrict} this
series is absolutely convergent:
\ba{interm2}
{\dis
 (0\vert e^{\psi'(\lambda)
f(\phi(\mu))} F([\phi],[\Phi]) \vert 0)}& = &
{\dis 
\frac{1}{2\pi i} \int\limits_{|z|=\rho} \,dz
\frac{e^{z^2\eta(\lambda,\lambda)/2} }{z-\alpha f\biggl(
z\xi(\lambda,\mu)\biggr)}}\non
&&\hspace{7mm}{\dis\times
 (0\vert e^{zp'_\psi(\lambda)}
  F([\phi],[\Phi]) \vert 0).}
\ea
Due to the relation \Eq{restrict} and the Rouch\'e theorem, the
equation
\be{equation}
z-\alpha f\biggl(z\xi(\lambda,\mu)\biggr)=0
\ee
has exactly one zero of the first order $z=z_0=z_0(\alpha)$ in the
domain $|z|<\rho$. Let us emphasize that \Eq{equation}
is a classical equation  where only complex functions are involved 
(i.e., no quantum operators). Thus, taking the integral with respect 
to $z$, we arrive at 
\be{scfin1} 
(0\vert 
e^{\left\{\alpha\psi'(\lambda) f(\phi(\mu))\right\}} F([\phi],[\Phi]) 
\vert 0)= \frac{
e^{z_0^2 \eta(\lambda,\lambda)/2} (0\vert e^{z_0 p'_\psi(\lambda)}
 F([\phi],[\Phi])\vert 0)}{
 1-\frac{\partial}{\partial z_0}
f(z_0\xi(\lambda,\mu))}
\ee
The right-hand side may be further simplified
\be{simplif}
e^{z_0^2 \eta(\lambda,\lambda)/2} (0\vert e^{z_0 p_\psi'(\lambda)}
 F([\phi],[\Phi])\vert 0)= \bra e^{z_0 \psi'(\lambda)} F([\phi],[\Phi]) \vert
0).
\ee

Let us recall once more that in the original formula \Eq{sc1} 
$f(\phi(\mu))$ was a quantum operator since it depends on the quantum 
field $\phi(\mu)$.  The result of the above calculation shows that 
this function may be replaced by a complex number $z_0$.  

It is worth mentioning, that one can analytically continue the result
obtained with respect to $\alpha$ into the domain where the inequality
\Eq{restrict} is not valid. We  propose the theorem,
\begin{thm}
\be{thm1}
 (0\vert e^{\left\{ 
 \alpha\psi'(\lambda)f\bigl(\phi(\mu)\bigr)\right\}} 
 F([\phi],[\Phi])\vert 0)=
\frac{\bra e^{z_0 \psi'(\lambda)} F([\phi],[\Phi]) \vert 0)}
{1-\alpha f'\Bigl(z_0\xi(\lambda,\mu)\Bigr)}.
\ee
Here the complex number $z_0$ can be found from the equation
\be{thm2}
z_0=\alpha f\Bigl(z_0\xi(\lambda,\mu)\Bigr),
\ee
and
\be{thm3}
f'\Bigl(z_0\xi(\lambda,\mu)\Bigr)=
\left.\frac{\partial}{\partial z}f\Bigl(z\xi(\lambda,\mu)\Bigr)
\right|_{z=z_0}\ .
\ee
\end{thm}

{\it Remark.}~~The equation \Eq{thm2} may have many solutions
if we do not impose the restriction \Eq{restrict}. In this
case one should choose the solution $z_0=z_0(\alpha)$, with
the property:
\be{prop}
\biggl.z_0(\alpha)\biggr|_{\alpha=0}=0.
\ee

Let us remind the reader that no operators are involved in 
equation \Eq{thm2} since $f(z)$ is a complex function. This is 
complex equation for the complex number $z_0$. 
\subsection{Matrix Case} 

The method of calculation of the mean value described above
can be easily generalized for more complicated case. Namely
let us consider the example:  
\be{mcmeanv} 
(0\vert 
\exp\left\{ \sum_{k=1}^N \psi'(\lambda_k) f_k(\phi(\lambda_k)) 
\right\} F([\phi],[\Phi]) \vert 0).  
\ee
It is clear that we may find the mean 
value in an analogous fashion to that for a scalar case.  Let us briefly
describe  the main steps of the corresponding derivation.

First, we have
\be{MCdec} 
 \exp\left\{ \sum_{k=1}^N \psi'(\lambda_k) f_k(\phi(\lambda_k)) 
\right\} = \prod_{j=1}^N 
\sum_{n_j=0}^\infty \frac{1}{n_j!} \bigl( 
\psi'(\lambda_j)\bigr)^{n_j} f_j^{n_j}\bigl(\phi(\lambda_j)\bigr).  
\ee
The normal ordering of $\psi'(\lambda_j)$ may be performed in a
similar fashion to (\ref{intrep}) and (\ref{sc3}):
\ba{mcinterm1}
&&{\dis\hspace{-2cm}
(0\vert \prod_{j=1}^N \left[\sum_{n_j=0}^\infty
 \frac{1}{n_j!} \bigl( \psi'(\lambda_j)\bigr)^{n_j}\right]  =
\frac{1}{(2\pi i)^N} \int \prod_{j=1}^{N}\frac{dz_j}{z_j^{n_j+1}}}\non
&&{\dis \hspace{25mm}\times
 \exp\left\{\frac{1}{2}\sum_{j,k=1}^N z_j z_k
\eta(\lambda_j,\lambda_k) \right\}
(0\vert e^{\sum_{k=1}^N z_k p'_\psi(\lambda_k)}.}
\ea
Here each of the integrals is taken by circle, which appears to be
a common domain of analyticity of the functions $f_k(z)$.

Next we find the mean value
\ba{mcinterm2}
&&{\dis
(0 \vert e^{\sum_{k=1}^N z_k p'_\psi(\lambda_k)}
 \prod_{k=1}^N f_k^{n_k}(\phi(\lambda_k)) ~ F([\phi],[\Phi])\vert 0)
 = \prod_{k=1}^N f_k^{n_k}\Bigl( \sum_{j=1}^N z_j
\xi(\lambda_j,\lambda_k)\Bigr)}\non
&&{\dis\hspace{4cm}
\times(0\vert e^{\sum_{k=1}^N z_k p'_\psi(\lambda_k)}
 F([\phi],[\Phi]) \vert 0).}
\ea
Now we substitute this into the expression for our mean value \Eq{mcmeanv}
and sum with respect to each $n_k$:
\ba{mcinterm3}
&&{\dis
(0\vert e^{\sum_{k=1}^N \psi'(\lambda_k) f_k(\phi(\lambda_k))}
F([\phi],[\Phi]) \vert 0)
= \frac{1}{(2\pi i)^N} \int \prod_{j=1}^{N}dz_j}\non
&&{\dis\times
\frac{e^{1/2 \sum_{j,k=1}^N z_j z_k \eta(\lambda_j,\lambda_k)}
(0\vert e^{\sum_{k=1}^N z_k p'_\psi(\lambda_k)} F([\phi],[\Phi])
\vert 0)}
{\prod_{j=1}^{N}\left[z_j- f_j\biggl(
\sum_{m=1}^N z_m \xi(\lambda_m,\lambda_j)\biggr)\right]}.}
\ea
In order to take the $z_j$ integral we introduce the
complex numbers $z_j^0$ as solutions of the system
\be{mcsystem}
z_j^0 = f_j\biggl( \sum_{m=1}^N z_m^0 \xi(\lambda_m,\lambda_j)\biggr).
\ee
We also define the matrix $M$:
\be{MCmatrix}
M_{jk} = \delta_{jk} -\left.{\partial\over\partial z_j}
 f_k\Bigl( \sum_{m=1}^N z_m \xi(\lambda_m,\lambda_k)\Bigr)
\right|_{z_l=z_l^0}.
\ee
After evaluation of the $z_j$ integration, we obtain the result
\ba{mcresult}
&&{\dis\hspace{-15mm}
(0\vert \exp\left\{ \sum_{k=1}^N \psi'(\lambda_k)
 f_k(\phi(\lambda_k)) \right\} F([\phi],[\Phi]) \vert 0)}\non
&&{\dis\hspace{2.5cm} =
\frac{ (0\vert \exp\left\{ \sum_{k=1}^N
\psi'(\lambda_k) z_k^0 \right\} F([\phi],[\Phi]) \vert 0)}
{ \det M}.}
\ea

As in scalar case, the system \Eq{mcsystem} may have many
solutions. In this case one can consider the replacement
$f_k\to \alpha f_k$. After this, it is necessary to choose the
solution of \Eq{mcsystem}, which  approaches zero as $\alpha
\to 0$ and to continue this solution to the point $\alpha=1$. 
\subsection{Continuous Case}

In order to evaluate the mean value \Eq{goal} let us consider the continuous
limit of \Eq{mcresult}:
\be{liml}
\lambda_{k+1}=\lambda_k+\Delta,
\ee
\be{limf}
f_k\bigl( \phi(\lambda_k)\bigr) = \Delta
  f\bigl(\lambda_k\vert \phi(\lambda_k)\bigr).
\ee
\be{limz}
z_k^0=\Delta z(\lambda_k),
\ee
and take the limit  $\Delta \rightarrow 0$.  We notice that
the system of constraints in (\ref{mcsystem}) turns into the integral
equation:
\be{CCinteq}
z(\lambda) = f\Bigl( \lambda\vert \int_{-\infty}^\infty d\mu~ z(\mu)
\xi(\mu,\lambda)\Bigr).
\ee
Furthermore, the matrix $M_{jk}$ becomes an integral operator
with the kernel
\be{CCkern}
M(\lambda,\mu) = \delta(\lambda-\mu) - {\delta\over\delta z(\mu)}
 f\Bigl( \lambda\vert \int_{-\infty}^\infty ds~ z(s) 
\xi(s,\lambda) \Bigr).
\ee
Equation (\ref{mcresult}) has the following continuous limit:
\ba{clresult}
&&{\dis\hspace{-12mm}
(0\vert \exp\left\{ \int_{-\infty}^\infty d\lambda~ \psi'(\lambda)
  f(\lambda\vert (\phi(\mu)) \right\} F([\phi],[\Phi]) \vert 0)}\non
&&{\dis\hspace{5mm}
 =(\det M)^{-1}
 (0\vert
 \exp\left\{ \int_{-\infty}^\infty 
d\lambda \psi'(\lambda) z(\lambda)\right\}
F([\phi],[\Phi]) \vert 0).}
\ea
Equation ~\Eq{CCinteq}--\Eq{clresult} are the main result of this
section. We would like to emphasize that the  commutativity of the quantum
fields \Eq{repdualfields} is extremely important for obtaining of 
this result.

The following calculations are trivial. We have
\ba{triv1}
&&{\dis\hspace{-15mm}
\bra \exp\left\{ \int_{-\infty}^\infty d\lambda
 \psi'(\lambda) z(\lambda)\right\}}
\non
&&{\dis\hspace{-10mm}=
\exp\left\{ \frac{1}{2}\int_{-\infty}^\infty 
d\lambda d\mu~\eta(\lambda,\mu)
 z(\lambda)z(\mu)\right\}
 \bra\exp\left\{ \int_{-\infty}^\infty d\lambda 
p'_\psi(\lambda) z(\lambda)\right\}.}
\ea
The action of the operator $\exp\{p'_\psi\}$ on the functional $F$ 
leads to the shift of the arguments of the last one (see \Eq{lm3}, 
\Eq{lm4}). We find
\ba{triv2}
&&{\dis
 (0\vert \exp\left\{ \int_{-\infty}^\infty d\lambda p'_\psi(\lambda) 
z(\lambda)\right\} F([\phi],[\Phi]) \vert 0)}\non
&&{\dis\hspace{5mm}=
F\left(\left[\int_{-\infty}^\infty\,d\lambda 
z(\lambda)\xi(\lambda,\mu)\right],
\left[\int_{-\infty}^\infty\,d\lambda 
z(\lambda)\tilde\xi(\lambda,\mu)\right]
\right).}
\ea
In the next section we shall use these results in order to evaluate the
mean value of the asymptotic expression \Eq{asy2}.
\section{The Mean Value of the Leading Term}

In order to find the mean value of the leading term of asymptotics
\Eq{asy2}, we need only to substitute the concrete expressions
\Eq{Af},  \Eq{AF} into the equations~\Eq{CCinteq}--\Eq{clresult} and
\Eq{triv2}.

An integral equation for the $z$-function is
\be{inteq}
z(\lambda)=-{i\over 2\pi} \sign(\lambda-\lambda_0) \ln\left\{
 1- \vartheta(\lambda) X(\lambda,\lambda_0) \right\},
\ee
where
\be{x}
X(\lambda,\lambda_0) \equiv 1+ \exp\left\{ \sign(\lambda-\lambda_0)
 \left( \ln{h(\lambda,\lambda_0)\over h(\lambda_0,\lambda)} +
\int_{-\infty}^\infty d\mu~{2ic z(\mu)\over c^2+(\lambda-\mu)^2} \right)
\right\}.
\ee
The integral operator $M$ has the kernel
\be{kern}
M(\lambda,\mu) = \delta(\lambda-\mu) -{i\over 2\pi} {\rm
sign}(\lambda-\lambda_0)
{\delta\over\delta z(\mu)} \ln\left\{ 1-\vartheta(\lambda)
 X(\lambda,\lambda_0) \right\}.
\ee
After evaluation of the variation derivative one should set $z(\lambda)$
equal to the solution of \Eq{inteq}.  

The functional $F([\phi],[\Phi])$ is given in \Eq{AF}.  It consists of
three factors: an  exponential factor, a power law correction with 
respect to $t$, and a constant factor, which depends only on the ratio 
$x/2t=\lambda_0$.  Using \Eq{triv2} and the integral equation 
\Eq{inteq}, we find  the exponential factor to be
\be{expfact} 
\exp\left\{ -i\int_{-\infty}^\infty d\lambda~
 ( x-2\lambda t)z(\lambda) \right\} .
\ee

The next factor is a power law correction
\be{nu1}
(2t)^{(\nu-1)^2/2},
\ee
where the expression for $\nu$ follows  from \Eq{nu}: 
\be{nu2}
\nu = \frac{i}{ 2\pi} \ln\left\{ \left[
1-\vartheta(\lambda_0)(1+e^{-\phi(\lambda_0)})\right] \left[
 1-\vartheta(\lambda_0)(1+e^{\phi(\lambda_0)})\right] \right\} .
 \ee
Instead of $\phi(\lambda_0)$ we substitute the integral
expression,
\be{phinu}
\phi(\lambda_0)\rightarrow u(\lambda_0)=\int_{-\infty}^\infty 
d\lambda~ \frac{2ic}{ c^2+(\lambda_0-\lambda)^2} z(\lambda), 
\ee
where the function $z(\lambda)$ is a solution of \Eq{inteq}. So
the power law correction becomes
\be{tnu1}
(2t)^{(\tilde\nu-1)^2/2},
\ee
where 
\be{tnu2}
\tilde\nu = \frac{i}{ 2\pi} \ln\left\{ \left[
1-\vartheta(\lambda_0)(1+e^{-u(\lambda_0)})\right] \left[
 1-\vartheta(\lambda_0)(1+e^{u(\lambda_0)})\right] \right\} .
 \ee
Finally, the constant term is equal to
\be{constfact}
g={\tilde C}([u(\mu)],[v(\mu)]) e^{v(\lambda_0)}
\exp\left\{ \frac{1}{2}\int_{-\infty}^\infty 
d\lambda d\mu~\eta(\lambda,\mu)
 z(\lambda)z(\mu)\right\},
\ee
where
\be{v}
v(\mu)=\int_{-\infty}^\infty 
d\lambda~ \frac{2(\lambda-\mu)}{ c^2+(\lambda_0-\lambda)^2} 
z(\lambda).
\ee

Thus we obtain for the mean value \Eq{asy2} of the operator $Q
(x,t)$
\ba{mainres}
&&{\dis\hspace{-15mm}
\bra Q(x,t)\ket =g\cdot e^{-it\lambda_0^2-ikt}
(2t)^{(\tilde\nu-1)^2/2}  (\det M)^{-1}}\non
&&{\dis\hspace{15mm}
\times
\exp\left\{ -i\int_{-\infty}^\infty d\lambda~
(x-2\lambda t) z(\lambda) \right\} .}
\ea
This is the main result of the paper.

Expression \Eq{mainres}  is the leading term of the asymptotic
evaluation of the Fredholm determinant representing the correlation
function \cite{ref2}.  In the next publication we shall study the
corrections.  The mean value of corrections  can contribute to  the
leading term of the asymptotic behavior of the correlation function.

\section*{ Acknowledgments}
We would like to thank Gordon Chalmers and Martin Bucher for useful
 discussion.
This work was  supported by the National Science Foundation (NSF)
Grant No. PHY-9321165, and the Russian Foundation of Basic Research
Grant No. 96-01-00344 and INTAS-93-1038. 



\begin{thebibliography}{99}
%
\bibitem{ref2}
T.~Kojima, V.~E.~Korepin and N.~A.~Slavnov, Determinant
representation of the Quantum nonlinear Schr\"odinger equation,
hep-th/9611216, (accepted to the Commun. Math. Phys.).
%
\bibitem{ref3}
T.~Kojima, V.~E.~Korepin and  N.~A.~Slavnov, Completely Integrable
equation for the quantum correlation function of nonlinear
Schr\"odinger equation, hep-th/9612255,
(accepted to the Commun. Math. Phys.).
%
\bibitem{ref4}
V.~E.~Korepin and  N.~A.~Slavnov, The Riemann-Hilbert
problem associated with quantum nonlinear Schr\"odinger equation.
Preprint ITP, SUNY at Stony Brook -97-33, hep-th/9706147, (submitted 
to Journal of Physics A.).
%
\bibitem{ref5} 
A.~R.~Its, V.~E.~Korepin 
and N.~A.~Slavnov, Asymptotical Solution of Riemann--Hilbert problem 
associated with quantum nonlinear Schr\"odinger equation. Preprint 
ITP, SUNY at Stony Brook-97-34. 
%
\bibitem{ref6} 
E.~H.~Lieb and  W.~Liniger, Phys.~Rev.~{\bf 130}, (1963), 1605.  
%
\bibitem{ref7} E.~H.~Lieb, Phys.~Rev.~{\bf 130}, (1963), 1616.  
%
\bibitem{ref8}
V.~E.~Zakharov and  A.~B.~Shabat, Sov.~Phys.~JETP {\bf 34}, (1972), 62.
%
\bibitem{ref9}
L.~D.~Faddeev and E.~K.~Sklyanin, Sov.~Phys.~Dokl.~{\bf 23}, (1978), 908.
%
\bibitem{ref10}
C.~N.~Yang and  C.~P.~Yang, J.~Math.~Phys.~{\bf 10}, (1969), 1115.
%
\bibitem{EFIK}
F.~H.~L.~E\ss{}ler, H.~Fhram, A.~R.~Its and V.~E.~Korepin,
J. Phys. A: Math. Gen. {\bf 29}, (1996), 5619.

\end{thebibliography}
\end{document}